\documentclass[twocolumn,showpacs,preprintnumbers]{revtex4}
\usepackage{graphicx}% Include figure files
\usepackage{dcolumn}% Align table columns on decimal point
\usepackage{bm}% bold math
\usepackage{amssymb}
\usepackage{epsfig}    
\usepackage{here}
\def\beq{\begin{equation}}
\def\eeq{\end{equation}}
\def\bea{\begin{array}}
\def\eea{\end{array}}
\def\be{\begin{equation}}
\def\ee{\end{equation}}
\def\ba{\begin{eqnarray}}
\def\ea{\end{eqnarray}}

\def\to{\rightarrow}

\def\dis{\displaystyle}
\def\f{\frac}
\def\[{\left[}
\def\]{\right]}
\def\({\left(}
\def\){\right)}

\def\sm0{{\widetilde{m}_0}}

\def\ts{\tilde{t}}
\def\cs{\tilde{c}}

\def\ov{\overline}

\def\U1em{{U(1)_{\rm em}}}
\def\to{\rightarrow}

\def\sq2{\sqrt{2}}

\def\tanb{\tan\,\beta}
\def\ee{e^+e^-}

\def\End{\end{document}}
\def\Journal#1#2#3#4{{#1} {\bf #2} (#4) #3}
\def\NPB{{\em Nucl. Phys.} B}
\def\PLB{{\em Phys. Lett.}  B}
\def\PRL{\em Phys. Rev. Lett.}
\def\PRD{{\em Phys. Rev.} D}

\def\EPC{{\em Euro. Phys. J.} C}

\begin{document}                                                              
%\draft

\title{Determining the Chirality of Yukawa Couplings \\
via Single Charged Higgs Boson Production \\                
       in Polarized Photon Collision
        }%
\author{%
{\sc Hong-Jian He\,$^1$,~~  Shinya Kanemura\,$^2$},~~    
{\sc C.--P. Yuan\,$^2$}
}
\affiliation{%
%\address{\vspace*{5mm}
\vspace*{2mm} 
$^1$Center for Particle Physics, 
University of Texas at Austin, Texas 78712, USA\\ 
$^2$Department of Physics and Astronomy, 
Michigan State University, East Lansing, Michigan 48824, USA
}
%\maketitle

%\vspace*{5mm} 
\begin{abstract}
\hspace*{-0.35cm}
When the charged Higgs boson is too heavy to be produced in pairs,
the predominant production mechanism at Linear Colliders 
is via the single charged Higgs 
boson production processes, such as  
$e^-e^+ \to b \bar c H^+, \, \tau \bar \nu H^+$
and 
$\gamma\gamma \to b \bar c H^+, \, \tau \bar \nu H^+$.
We show that the yield of a heavy charged Higgs boson 
at a $\gamma\gamma$ collider is typically one or two orders
of magnitude larger than that at an  $e^-e^+$ collider.
Furthermore, a polarized $\gamma\gamma$ collider can 
determine the chirality of the Yukawa couplings of fermions 
with charged Higgs boson via single charged Higgs 
boson production, and thus discriminate models of new physics. 
\pacs{\,12.60.-i,\,12.15.-y,\,11.15.Ex 
\hfill   ~~ [ March, 2002 ~and~ hep-ph/0203090 ] }

\end{abstract}
%\vskip 1pc]

\maketitle

\setcounter{footnote}{0}
\renewcommand{\thefootnote}{\arabic{footnote}}

\section{Introduction} 

The detection of a charged Higgs boson unambiguously signals new physics 
beyond the Standard Model (SM) of particle physics.
In most extensions of the SM, the mass ($M_{H^\pm}$) of 
a charged Higgs boson ($H^\pm$) is predicted to be
 around the weak scale. 
At hadron colliders, such as the Fermilab Tevatron and the  
CERN Large Hadron Collider (LHC), a light charged Higgs boson 
can be produced from the decay of top quark via 
$\, t \to H^+ b$, if $M_{H^\pm} < m_t-m_b$, where
$m_t$ and $m_b$ are the masses of top and bottom quarks,
respectively~\cite{tbH}.
Otherwise, it can be produced in pairs via the $s$-channel 
$q \bar q$ fusion process~\cite{ppHpHm} through the gauge interactions of 
$\gamma$-$H^+$-$H^-$ and $Z$-$H^+$-$H^-$. 
However, the rate of the pair production is generally 
smaller than that predicted by  
the single charged Higgs boson production mechanisms  
as the mass of the charged Higgs boson increases. 
At hadron colliders, 
a single charged Higgs boson can be produced via 
$g b \to H^\pm t$ \cite{gbHt}, 
$cs, cb \to H^\pm$ \cite{dhy,hy}, and 
$gg$ (or $q\bar q$) $\to H^\pm W^\mp$\cite{ppHW}, etc.  
While the single charged Higgs boson production rate at hadron 
colliders can be sizable, it is extremely difficult to determine
the chirality of the Yukawa couplings of fermions with charged
Higgs boson.
In this work, we show that it can be readily achieved 
at a polarized photon collider.

If $M_{H^\pm}$ is smaller than half of the center-of-mass
energy ($\sqrt{s}$) of a Linear Collider (LC), then $H^\pm$ can be 
copiously produced in pairs via the scattering processes
$e^-e^+ \to H^- H^+$ and $\gamma \gamma \to H^- H^+$~\cite{eeHpHm}.
The production rate of a $H^- H^+$ pair is determined by 
the electroweak gauge interaction of $H^\pm$ and gauge bosons, which 
depends only on the electric charge and weak-isospin of $H^\pm$.
When $M_{H^\pm} > \sqrt{s}/2$, it is no longer possible to produce the 
charged Higgs bosons in pairs. In this case, the predominant production
mechanism of the charged Higgs boson is via the 
single charged Higgs boson production processes, such as 
the loop induced process 
$e^-e^+ \to H^\pm W^\mp$~\cite{eeWH}, 
and the tree level processes 
$e^-e^+ \to b \bar c H^+, \, \tau \bar \nu H^+$
and 
$\gamma\gamma \to b \bar c H^+, \, \tau \bar \nu H^+$.
The production rate of the above tree level processes depends on the 
Yukawa couplings of fermions with $H^\pm$.
This makes it possible to discriminate models of flavor 
symmetry breaking by measuring the production rate of the single 
charged Higgs boson at the LC.
We find that the yield of a heavy charged Higgs boson 
at a $\gamma\gamma$ collider is typically one or two orders
of magnitude larger than that at an $e^-e^+$ collider.
[In this paper, we assume the center-of-mass energy of 
a $\gamma\gamma$ collider ($\sqrt{s}_{\gamma\gamma}$) 
is about 80\% of that of an $e^-e^+$ collider ($\sqrt{s}_{ee}$).]

It is well known that the main motivation for building a
 high-energy polarized photon collider is 
 to determine the $CP$ property of 
the neutral Higgs bosons \cite{CPprop}.  
In this study, we provide another motivation 
for having a polarized photon collider -- 
to determine the chirality of the Yukawa coupling of 
fermions with charged Higgs boson via single charged Higgs 
boson production so as to discriminate models of flavor symmetry
breaking.

\vspace*{-3mm}
\section{Yukawa Interactions in New Physics Models}
The well-motivated new physics models that contain a charged 
Higgs boson can be classified into 
the weakly interacting models 
(with elementary Higgs scalars) \cite{Kane}
and the strongly interacting models
(with composite Higgs scalars) \cite{sekhar}.
A typical example of the weakly interacting models is the 
Minimal Supersymmetric SM (MSSM) \cite{MSSM}, and
of the strongly interacting models is  
the dynamical Top-color (TopC) model \cite{Hill}.
In general, the Yukawa coupling of the fermions with $H^\pm$ 
can be written as
\beq
\label{eq:yukawa}
{\cal L}_{\rm Y} = \ov{f'} 
\left( Y_L^{f'f} P_{L} + Y_R^{f'f} P_{R} \right)
       f \,H^- +{\rm h.c.}\,, 
\eeq 
where $f$ and $f'$ represent up-type and down-type fermions, respectively, 
and  
$P_{L,R}$ are the chirality projection operators 
$ P_{L,R}= \(1\mp \gamma_5\)/2$\,.

We first consider the Yukawa sector of the MSSM, which is similar to
that of a Type-II two-Higgs-doublet model (2HDM).
The corresponding tree-level Yukawa couplings of fermions 
with $H^\pm$ are given by 
\begin{eqnarray}
\label{eq:yukawaC}
Y_{L(0)}^{f'f} \!=\! \frac{\sqrt{2} m_{f'}}{v} V_{ff'}  \tan\!\beta, \;\;\;
Y_{R(0)}^{f'f} \!=\! \frac{\sqrt{2} m_f}{v} V_{ff'}  \cot\!\beta,  
\end{eqnarray}
where $m_{f}$ ($m_{f'}$) is  the mass of fermion 
$f$ ($f'$), 
$\tan\beta = \langle H_u\rangle / \langle H_d\rangle$
is the ratio of the vacuum expectation values of 
two Higgs doublets,
$v = ({\langle H_u\rangle}^2 + {\langle H_d\rangle}^2)^{1/2} 
  \simeq 246$\,GeV,
and $V_{ff'}$ is the relevant 
Cabibbo-Kobayashi-Maskawa (CKM) matrix element. 
The coupling constants 
$Y_{L(0)}^{f'f}$ and $Y_{R(0)}^{f'f}$ 
vary as the input parameter $\tan\beta$ changes.   
For the $\bar \tau$-$\nu$-$H^-$ coupling,  
$Y_{L(0)}^{\tau\nu}$ increases as $\tan\beta$ grows, and 
reaches about $0.20-0.51$ for $\tan\beta=20-50$, while 
$Y_{R(0)}^{\tau\nu}$ is zero because of the absence of
right-handed Dirac neutrinos in the MSSM. 
As a typical choice for large $\tanb$, we take
\beq
\(Y_{L(0)}^{\tau\nu},\, Y_{R(0)}^{\tau\nu}\)
\,\simeq\, (0.3,~0)\,,
~~~~{\rm for}~ \tanb \simeq 30\,.
\label{eq:taunuH-MSSM-def}
\eeq                    
The tree level $\bar b$-$c$-$H^-$ coupling 
contains a CKM suppression factor $V_{cb}\simeq 0.04$, so that 
$Y_{L(0)}^{bc}$ is around $0.03$ for $\tan\beta=50$ and  
$Y_{R(0)}^{bc}$ is less than about $2\times 10^{-4}$ for $\tan\beta > 2$. 
However, some radiative corrections arising from the
Supersymmetry (SUSY) interaction can 
significantly enhance 
the tree level $\bar b$-$c$-$H^-$ coupling.
It was shown in Ref.\,\cite{dhy} that 
the radiatively generated $\bar b$-$c$-$H^-$ coupling from 
the natural stop-scharm ($\ts-\cs$) mixing 
in the SUSY soft-breaking sector
can be quite sizable.
For instance, in some class of MSSM~\cite{dhy}, the 
non-diagonal scalar trilinear $A$-term for the up-type squarks 
can induce the Yukawa couplings of $b$-$c$-$H^\pm$, such that 
for $\tan\beta=50$,  $Y_L^{bc}$ is around 0.05 and 
$Y_R^{bc}$ is about 0.
The feature that 
 $Y_L^{bc} \gg  Y_R^{bc} \sim 0$ 
will be relevant to our latter discussion on 
how to use a polarized photon collider to discriminate 
models of new physics by testing the chirality structure 
of the fermion Yukawa couplings.

We now turn to the TopC model \cite{Hill},
which provides an attractive scenario for explaining the large
top quark mass from the $\langle \bar{t}t \rangle$ condensation
via the strong $SU(3)_{\rm tc}$ TopC interaction at the TeV scale.
The associated strong tilting $U(1)$ force is attractive in the
$\langle \bar{t}t \rangle$ channel and repulsive in the
$\langle \bar{b}b \rangle$ channel, so that the bottom quark
mainly acquires its mass from the TopC instanton contribution
\cite{Hill}.
This model predicts three relatively light physical 
top-pions $(\pi_t^0,\,\pi^\pm_t )$.
The Yukawa interactions of these top-pions 
with the $b$, $t$ and $c$ (charm) quarks can be written as 
\beq
\bea{l}
\dis\f{m_t\tan\beta'}{v}\hspace*{-1.1mm}\left[
i{K_{UR}^{tt}}
{K_{UL}^{tt}}^{\hspace*{-1.3mm}\ast}\overline{t_L}t_R\pi_t^0
\hspace*{-0.7mm}+\hspace*{-0.8mm}\sq2
{K_{UR}^{tt}}{K_{DL}^{bb}}^{\hspace*{-1.3mm}\ast}
\overline{b_L}t_R\pi_t^- 
+ \right.
\\[3.3mm]
~~\left.
i{K_{UR}^{tc}}
 {K_{UL}^{tt}}^{\hspace*{-1.3mm}\ast}\overline{t_L}c_R\pi_t^0
\hspace*{-0.7mm}+\hspace*{-0.8mm}\sq2
{K_{UR}^{tc}} {K_{DL}^{bb}}^{\hspace*{-1.3mm}\ast}
\overline{b_L}c_R\pi_t^-
\hspace*{-0.7mm}+\hspace*{-0.7mm}{\rm h.c.}  \right],
\eea
\label{eq:Ltoppi}
\eeq
where {\small $\tan\beta' = \sqrt{(v/v_t)^2-1}$} and
the top-pion decay constant
$v_t\simeq O(60-100)$~GeV.
The rotation matrices $K_{UL,R}$ and $K_{DL,R}$ are needed 
for diagonalizing the up- and down-quark mass matrices
$M_U$ and $M_D$, i.e.,
{\small $~K_{UL}^\dag M_U K_{UR} = M_U^{\rm dia}~$} and
{\small $~K_{DL}^\dag M_D K_{DR} = M_D^{\rm dia}$},~
from which the CKM matrix is defined as
\,{\small $V=K_{UL}^\dag K_{DL}$}\,.\,  
To yield a realistic form of 
the CKM matrix $V$ (such as
the Wolfenstein-parametrization), the TopC model 
generally predicts
$K_{UR}^{tt}\simeq 0.99\hspace*{-0.5mm}-\hspace*{-0.5mm}0.94$, 
$K_{UR}^{tc}\lesssim 0.11\hspace*{-0.5mm}-\hspace*{-0.5mm}0.33$, and
$K_{UL}^{tt} \simeq K_{DL}^{bb} \simeq 1$,
which suggests that the $t_R$-$c_R$ transition
can be naturally around $10-30\%$~\cite{hy}. 
Consequently, the Yukawa couplings of fermions with
the charged top-pion (pseudoscalar Higgs boson) are   
$Y_L^{bt} \,=\, Y_L^{bc} \,=\, 0\,$,           
$Y_R^{bt} \,\simeq \, (\sqrt{2}m_t/v)\tan\beta' $, and
$Y_R^{bc} \,\simeq \, Y_R^{bt} K_{UR}^{tc}$. 
Thus, taking a typical value of $\tan\beta' \simeq 3$ and 
a conservative input of $ K_{UR}^{tc} \simeq 0.1 $
for the \,$t_R-c_R$\, mixing,  we obtain   
\beq
Y_R^{bt} \simeq 3\,, 
~~~~{\rm and}~~~~ 
\(Y_L^{bc},\,Y_R^{bc}\) 
\simeq (0,\, 0.3) \,,
\label{eq:bcH-TopC-def}
\eeq
which will be used as the sample 
TopC parameters for our numerical analysis.

As noted earlier, the chirality of the Yukawa couplings of
$b$-$c$-$\pi_t^\pm$ is purely 
 right-handed, in contrast to the left-handed 
 Yukawa couplings predicted by the MSSM with a large $\tan\beta$ value.
This makes it possible to discriminate 
the TopC model from the MSSM or a Type-II 2HDM 
by measuring the production rate of single charged Higgs 
boson at a polarized photon collider.
\begin{figure}
\includegraphics[width=8.1cm,height=6cm]{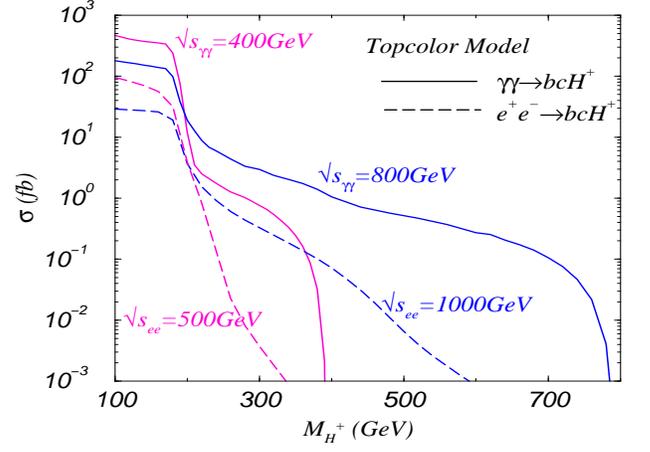}
%{\par\centering \resizebox*{0.4\textwidth}
%{!}{\includegraphics{aa.vs.ee_bch.eps}} \par}
\vspace*{-2mm}
\caption{Cross sections of $\gamma\gamma\to b \bar c H^+$ (solid curves) 
         and $e^+e^-\to b \bar c H^+$ (dashed curves) for the TopC model
[cf. Eq.\,(\ref{eq:bcH-TopC-def})]
         with unpolarized photon beams  
         at $\sqrt{s}_{\gamma\gamma}=400$ ($800$) GeV 
            and $\sqrt{s}_{ee}=500$ ($1000$) GeV. 
}
\label{fig:bch_topc_tot}
\end{figure}

\vspace{-3mm}
\section{
$H^\pm$ Production in Photon Collision as
a Probe of New Physics}
Using the default parameters of the TopC model, as 
described in the previous section, 
cf. Eq.~(\ref{eq:bcH-TopC-def}), we calculate 
the total cross sections of 
$\gamma\gamma \to b \bar c H^+$ and 
$e^+e^- \to b \bar c H^+$ as a function of $M_{H^\pm}$.
The result  is shown in Fig.~\ref{fig:bch_topc_tot}.
A few discussions on the feature of the results shown in 
Fig.~\ref{fig:bch_topc_tot} are in order.
For $M_{H^\pm} < \sqrt{s}/2$, 
the charged Higgs bosons can be produced in pairs.
In this case, the production cross section for 
$\gamma\gamma \to b \bar c H^+$ 
(and  $e^-e^+ \to b \bar c H^+$) is dominated by the contribution 
from the $H^-H^+$ 
pair production diagrams with the produced $H^-$ decaying 
into a $b \bar c$ pair. 
Hence, its rate is proportional to the decay branching ratio
$Br(H^- \to b {\bar c})$.
(For example, for a 200\,GeV (400\,GeV) $H^\pm$ , 
$\Gamma_H$  is about 7\,GeV (143\,GeV), and 
$Br(H^- \to b {\bar c})$ is 0.15 (0.015).)
As shown in the figure, there is a {\it kink} structure when 
$M_{H^\pm}$ is around 180\,GeV. That is caused by the change in 
 $Br(H^- \to b {\bar c})$ when the decay channel 
 $H^- \to b {\bar t}$ becomes available.
Furthermore, for $M_{H^\pm} < \sqrt{s}_{\gamma\gamma}/2$, the cross section 
in $\gamma\gamma$ collision is 
typically an order of magnitude larger 
than that in $e^-e^+$ collision.

It is evident that the cross section of
 $\gamma\gamma \to b \bar c H^+$ 
is larger than that of $e^+e^- \to b \bar c H^+$ in 
the whole $M_{H^\pm}$ region. 
For $M_{H^\pm} > \sqrt{s}/2$, where 
the pair production is not kinematically allowed,
 the difference between these two cross sections becomes much larger 
 (two or three orders of magnitude) for a larger 
$M_{H^\pm}$ value.  
To understand the cause of this difference, we have to examine
the Feynman diagrams that contribute to the scattering processes
$e^-e^+ \to b \bar c H^+$
and
$\gamma\gamma \to b \bar c H^+$ .
In the former process, all the Feynman diagrams contain an $s$-channel 
propagator which is either a virtual photon or a virtual $Z$ boson.
Therefore, when $M_{H^\pm}$ increases for a fixed 
$\sqrt{s}_{ee}$, the 
cross section decreases rapidly.
On the contrary, in the latter process, when 
$M_{H^\pm} > \sqrt{s}_{\gamma\gamma}/2$,
the dominant contribution arises from the fusion diagram 
$\gamma\gamma \to (c \bar c) (b \bar b) 
\to  b \bar c H^+$, whose contribution is enhanced by the 
two collinear poles (in a $t$-channel diagram) 
generated from  $\gamma \to c \bar c$ and 
$\gamma \to b \bar b$ in high energy collisions.
Since the collinear enhancement takes the form of $\ln (M_{H^+}/m_q)$, 
with $m_q$ being the bottom or charm quark mass, 
the cross section of $\gamma\gamma \to b \bar c H^+$ 
does not vary much as $M_{H^\pm}$ increases until it is close
to $\sqrt{s}_{\gamma\gamma}$.

\begin{figure}
\includegraphics[width=8.1cm,height=5.4cm]{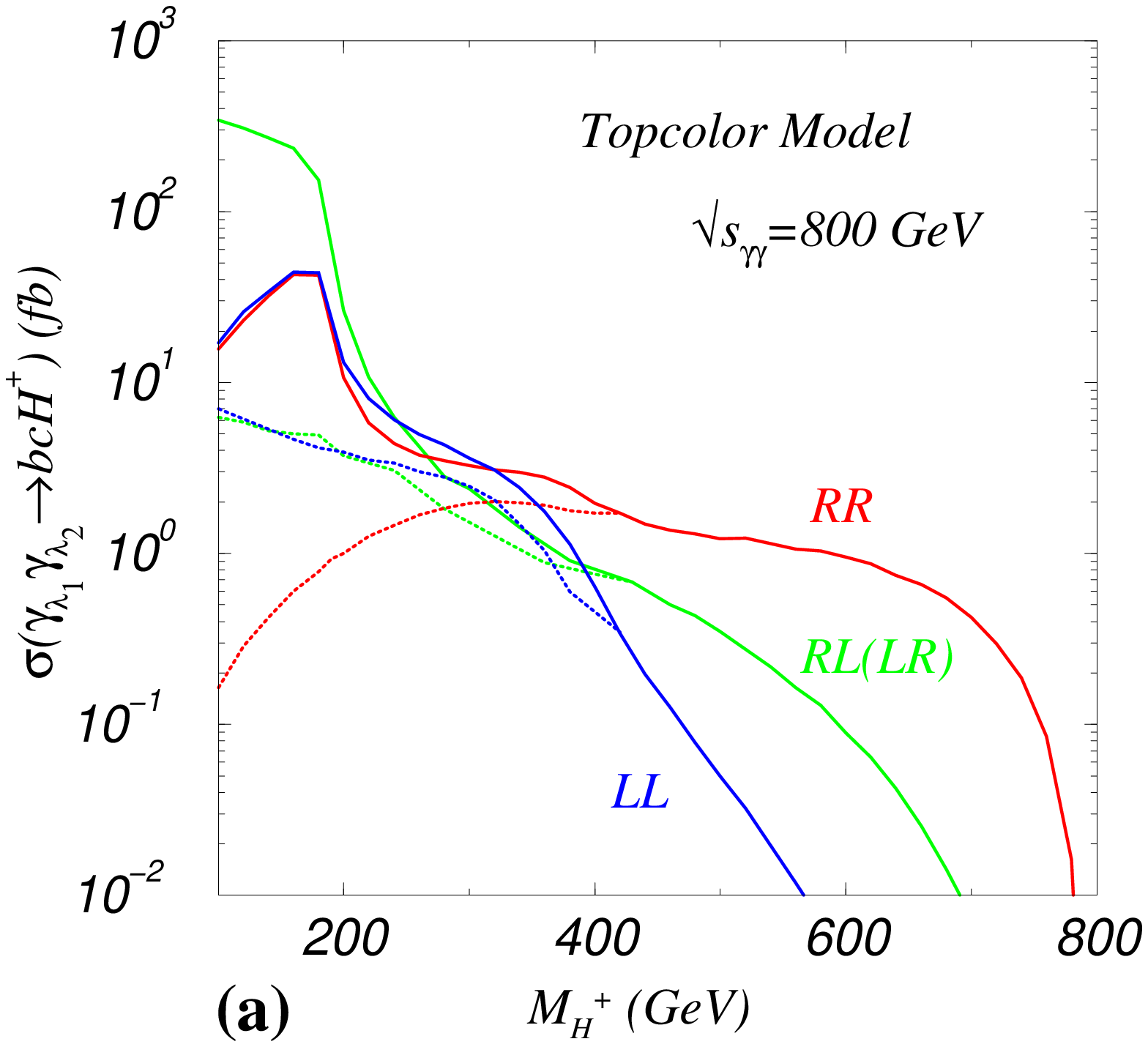} 
\\
\includegraphics[width=8.1cm,height=5.4cm]{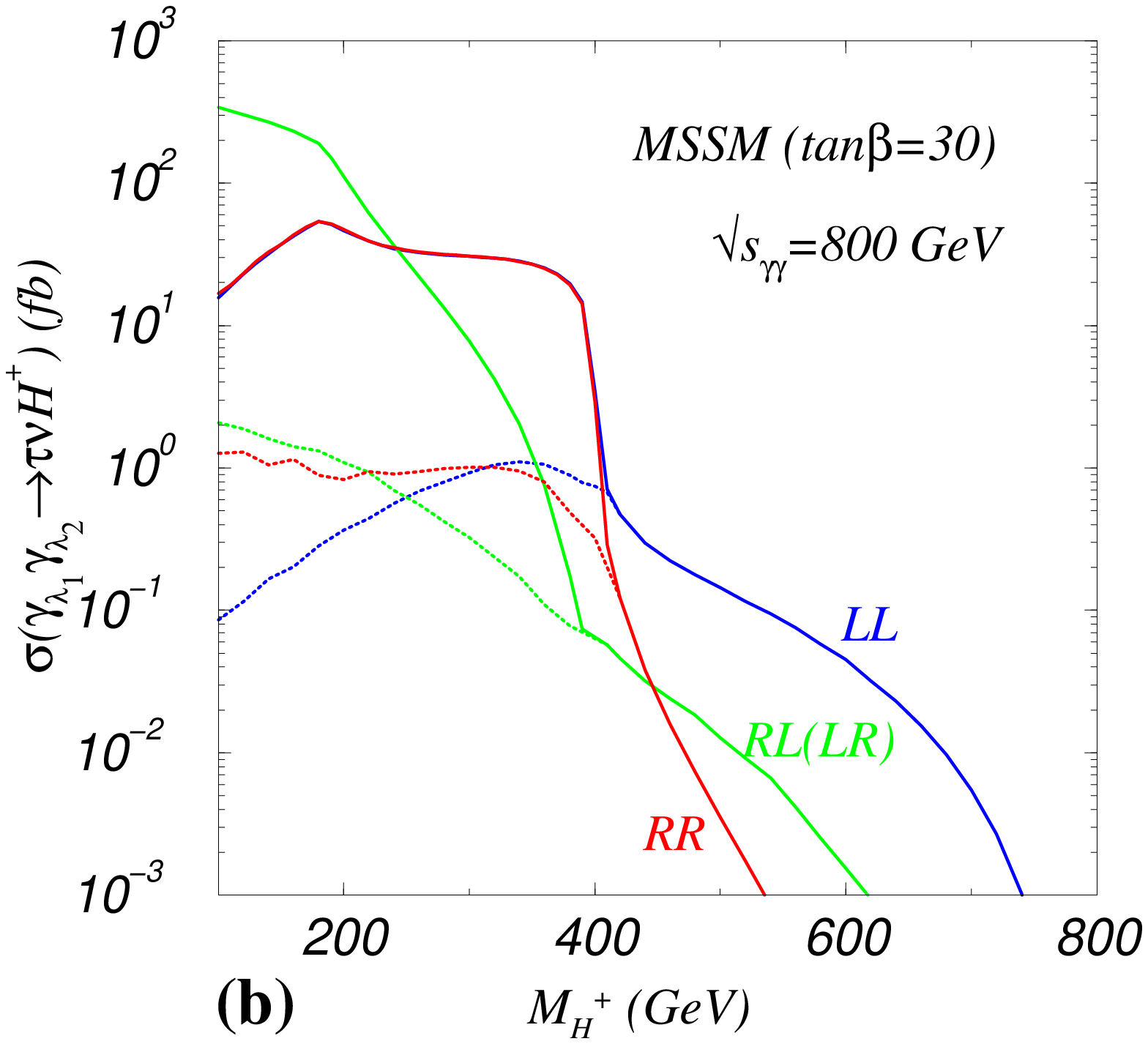}
\vspace*{-2mm}
\caption{(a): Cross sections of 
$\gamma_{\lambda_1}\gamma_{\lambda_2} \to b {\bar c} H^+$ 
         at $\sqrt{s}_{\gamma\gamma}=800$ GeV in  
polarized photon collisions for the TopC model            
[cf. Eq.\,(\ref{eq:bcH-TopC-def})].         
Solid curves are the results without any kinematical cut, and 
Dashed curves are the results with the kinematical cuts 
specified in the text [cf. Eq.\,(\ref{eq:kin-cut})]. 
(b): 
Cross sections of 
$\gamma_{\lambda_1}\gamma_{\lambda_2} \to \tau\nu H^+$ 
at $\sqrt{s}_{\gamma\gamma}=800$ GeV 
in polarized photon collisions for the 
MSSM [cf. Eq.\,(\ref{eq:taunuH-MSSM-def})]. 
}
\label{fig:bch_topc_pol}
\end{figure}

From the above discussions we conclude that a photon-photon collider is 
superior to an electron-positron collider to detect a 
heavy charged Higgs boson.  
Moreover, a polarized photon collider can determine the 
chirality of the Yukawa couplings of fermions with charged Higgs 
bosons via single charged Higgs boson production. 
This point is illustrated as follows.
First, let us consider the case that 
$M_{H^\pm} > \sqrt{s}_{\gamma\gamma}/2$.
As noted above, in this case, the production 
cross section is dominated by the fusion 
diagram $\gamma\gamma \to (c \bar c) (b \bar b) 
\to  b \bar c H^+$. In the TopC model, because 
$Y_L^{bc}=0$ (and $Y_R^{bc} \neq 0$), it corresponds to   
$\gamma\gamma \to (c_R \bar c_R) (b_L \bar b_L) 
\to  b_L \bar c_R H^+$. 
On the other hand, in the MSSM with stop-scharm mixing and 
large $\tanb$, $Y_R^{bc} \sim 0$ (and $Y_L^{bc} \neq 0$), 
it becomes 
$\gamma\gamma \to (c_L \bar c_L) (b_R \bar b_R) 
\to  b_R \bar c_L H^+$. 
Therefore, we expect that if both photon beams are 
right-handedly polarized (i.e. $\gamma_R \gamma_R$), then
a TopC charged Higgs boson (top-pion) can be 
copiously produced, while a MSSM charged Higgs boson 
(with a large $\tan\beta$) is highly suppressed.
To detect a MSSM charged Higgs boson, both photon beams 
have to be left-handedly polarized (i.e. $\gamma_L \gamma_L$).
This is supported by an exact calculation whose results
are shown in Fig.~\ref{fig:bch_topc_pol}(a) for the TopC model.
(A similar feature also holds for the class of 
MSSM proposed in Ref.\,\cite{dhy}  
after interchanging 
the label of $RR$ and $LL$ in Fig.~\ref{fig:bch_topc_pol}(a).)

The feature of the polarized photon cross sections 
for $M_{H^\pm} < \sqrt{s}/2$ can be understood from examining
the production process $\gamma\gamma \to H^+H^-$. 
The helicity amplitudes for the  $H^+H^-$ pair production   
in polarized photon collisions are calculated as 
\begin{eqnarray}
\label{eq:pair}
&& M(\gamma_{\lambda_1}^{} \gamma_{\lambda_2}^{} \to H^+H^-) = 
\nonumber\\
&&   2 e^2  \lambda_1 \lambda_2   
\left( \frac{1 - \xi^2}{1 - \xi^2 \cos^2 \Theta } \right) 
  +  e^2 \left( 1 - \lambda_1 \lambda_2 \right),      
\end{eqnarray}
where the degree of polarization of the initial state photons, 
$\lambda_{1}$ and $\lambda_{2}$, can take 
the value of either $-1$ or $+1$, 
corresponding to a left-handedly ($L$) or right-handedly ($R$) 
polarized photon beam, respectively;  
$\Theta$ is the scattering angle of $H^+$ in the center-of-mass
frame; and  $ \xi=\sqrt{ 1 - 4 M_{H^\pm}^2 / s_{\gamma\gamma} }$.
In the massless limit, i.e. when $M_{H^\pm} \to 0$, the above result 
reduces to  
$M(\gamma_{\lambda_1}^{}\gamma_{\lambda_2}^{}\to H^+H^-) \simeq 
 e^2 \left( 1 - \lambda_1 \lambda_2 \right)$. 
Denote 
$\sigma_{ \lambda_{1} \lambda_{2}}^{\rm pair}$ as the cross section 
of $\gamma_{\lambda_1}^{}\gamma_{\lambda_2}^{}\to H^+H^-$.
We find that $\sigma_{LR}^{\rm pair}=\sigma_{RL}^{\rm pair}$, and they
dominate the total cross section when 
$ M_{H^\pm}^2 \ll s$, while 
$\sigma^{\rm pair}_{LL}$ and $\sigma_{RR}^{\rm pair}$ 
are equal and approach to zero as $M_{H^\pm} \to 0$.
Since for $M_{H^\pm} < \sqrt{s}_{\gamma\gamma}/2$ the bulk part of the 
cross section of $\gamma\gamma \to b \bar c H^+$ 
comes from  
$\sigma(\gamma\gamma\to H^+H^-) \times Br(H^- \to b \bar c )$,
the $LL$ and $RR$ cross sections are smaller than the
$LR$ ($=RL$) cross sections as $M_{H^\pm}$ 
decreases, cf. Fig.~\ref{fig:bch_topc_pol}.

It is important to point out that the complete set of 
Feynman diagrams has to be included to calculate 
$\sigma(\gamma\gamma \to b \bar c H^+)$ even when 
 $M_{H^\pm} < \sqrt{s}_{\gamma\gamma}/2$ 
because of the requirement of gauge invariance.
To study the effect of the additional Feynman diagrams, other
than those contributing to the $H^+H^-$ pair production from 
$\gamma\gamma\to H^+H^-(\to b \bar c )$,
one can examine the {\it single} charged Higgs boson rate 
in this regime with the requirement that the invariant mass of 
$b \bar c$, denoted as $M_{b\bar c}$, satisfies 
the following condition:
\begin{eqnarray}
\label{eq:kin-cut}
|M_{b \bar c} & - & M_{H^\pm}|  >   \Delta M_{b\bar c} \, ,
\qquad {\rm with} \nonumber\\
\Delta M_{b\bar c} & = &
\min \left[  25\,{\rm GeV},\,
\max\left[ 1.18 M_{c\bar b} {2 \delta m \over m}, \Gamma_H \right] 
\right] \, ,
\nonumber\\ 
\qquad 
{\delta m \over m} & = & {0.5 \over \sqrt{M_{b\bar c}/2} }
\, ,
\end{eqnarray}
where ${\delta m \over m}$ denotes the mass resolution of 
the detector for observing the final state $b$ and ${\bar c}$ jets 
originated from the   %\linebreak
\noindent
decay of $H^-$.  
For example, in Fig.~\ref{fig:bch_topc_pol}(a) the
set of dashed-lines are the polarized cross sections after 
imposing the above kinematical cut.
With this cut, the total rate reduces by about one order of 
magnitude for $M_{H^\pm} < \sqrt{s}_{\gamma\gamma}/2$. 
(Needless to say that this cut can hardly change 
the event rate when $M_{H^\pm} > \sqrt{s}_{\gamma\gamma}/2$.)
The effect of this kinematic cut on the 
$RR$ and $LL$ rates are significantly different in the 
low $M_{H^\pm}$ region. It implies that the $H^+H^-$ pair 
production diagrams cannot be the whole production mechanism, 
otherwise,
we would expect the rates of $RR$ and $LL$ be always equal 
due to the parity invariance of the QED theory.
(Again, a similar feature also holds for the 
MSSM models proposed in Ref.~\cite{dhy} after 
interchanging the labels of $LL$ and $RR$.)
In Fig.~\ref{fig:bch_topc_pol}(b), we show the results
for $\gamma_{\lambda_1}\gamma_{\lambda_2} \to \tau\nu H^+$
in the MSSM with $\tanb=30$.

\vspace*{-5mm}
\section{Conclusions}
\vspace*{-2mm}
The single charged Higgs boson can be produced at polarized
photon colliders via the scattering processes 
$\gamma \gamma \to b \bar c H^+$ and 
$\gamma \gamma \to \tau \nu H^+$. 
We have shown that  
the production rate of $H^\pm$ in the $\gamma\gamma$ collision 
is much larger than that in the $e^-e^+$ collision, cf.
Fig.~\ref{fig:bch_topc_tot}.
For $M_{H^+}> \sqrt{s}/2$, 
the $e^+e^-$ rate is smaller by at least one 
or two orders of magnitude than the $\gamma \gamma$ rate.
This is because in high energy collisions there 
are two collinear poles 
($\gamma\gamma \to (c \bar c) (b \bar b) \to  b \bar c H^+$)  
in
$\gamma \gamma \to b \bar c H^+$, while  
the $e^+e^-$ processes contain only $s$-channel diagrams 
and cannot generate any collinear enhancement factor to  
the single charged Higgs boson production rate.
Furthermore, we find that it is possible to measure 
the Yukawa couplings $Y_L$ and $Y_R$, separately, 
at photon-photon colliders
by properly choosing the polarization states of the 
incoming photon beams.
This unique feature of the photon colliders can 
be used to discriminate models of flavor symmetry breaking.
Hence, we conclude that a polarized photon-photon collider is not only 
useful for determining the CP property of a neutral Higgs boson, 
but also useful for detecting a heavy charged Higgs boson and 
determining the chirality of the Yukawa couplings of fermions with 
charged Higgs boson.

Needless to say that a full Monte Carlo simulation is needed to determine 
how well the Yukawa couplings can be measured. For detecting a heavy 
charged Higgs boson (say, 500\,GeV in the TopC model),
the SM background rates can be ignored when $H^+$ predominantly decays
into a $t {\bar b}$ pair, and the signal rate is at most reduced by a 
factor of 2 to 3 by demanding the polar angle ($\theta$) of the forward $b$ 
and $\bar c$ jets to satisfy $|\cos \theta| < 0.98$ or $0.95$,
respectively, at a 800\,GeV photon-photon collider.

\vspace*{3mm}
\noindent
{\bf Acknowledgments}~~~
We thank G.L. Kane for valuable discussions on the
SUSY flavor mixing.
SK would like to thank S. Moretti for useful discussions 
and for comparing part of our results with his calculation. 
This work was supported in part by the NSF grant PHY-0100677
and DOE grant DEFG0393ER40757.

%\end{narrowtext}
\end{document}